\documentclass[twocolumn,amsmath,amssymb]{revtex4}
\begin{document}
\title{External fields, density functionals, and the Gibbs inequality}
\author{John D. Weeks}
%\email{jdw@ipst.umd.edu}
\affiliation{%
Institute for Physical Science and Technology\\
and Department of Chemistry and Biochemistry,\\
University of Maryland, College Park, MD 20742
}
%\date{\today}

\begin{abstract}
By combining the upper and lower bounds to the free energy as given by the
Gibbs inequality for two systems with the same intermolecular interactions
but with external fields differing from each other only in a finite region
of space $\Gamma $, we show that the corresponding equilibrium densities
must also differ from each other somewhere in $\Gamma .$ We note that the
basic equations of density functional theory arise naturally
from a simple rearrangement and reinterpretation of the terms in the upper
bound Gibbs inequality for such systems and briefly discuss some of the
complications that
occur when the intermolecular interactions of the two systems also differ.
\end{abstract}

\maketitle

\section{ Introduction}
Michael Fisher's work has often shown us that important insights can arise
from simple starting points through the use of basic principles of
statistical mechanics. With this in mind, (but on a much more modest
scale!), we will examine here some consequences of one of the earliest
fundamental relations of statistical mechanics: the Gibbs inequality,
through which Gibbs established the basic convexity properties of the free
energy \cite{1}. The usefulness of this inequality in the development of
density functional theory (both classical \cite{2,3,4,5} and quantum
versions \cite{6,7}) and in perturbation theories of liquids \cite{4,8,9}
has long been recognized \cite{10}. We report here some very simple
consequences of the Gibbs inequality for the density response of a fluid to
an external field that do not appear to be widely known, and show that the
basic equations of density functional theory arise directly from a
rearrangement and reinterpretation of terms in a special case of the Gibbs
inequality.

The Gibbs inequality in its most general form relates the equilibrium
properties of a system of interest with Hamiltonian $H(\Phi )$ (with general
intermolecular potentials and a generalized single particle potential
$\Phi $ simply
related to the external field $\phi $, as discussed below) to those of a
``trial'' system with Hamiltonian $H_{t}(\Phi _{t})$ with possibly different
intermolecular potentials and field $\Phi _{t}.$ We consider here the Grand
Canonical ensemble, where the free energy $\Omega =-k_{B}T\ln \Xi $ and
density distribution functions such as the singlet density $\rho (\mathbf{r})
$ are functions of the temperature $T,$ volume $V$ and chemical potential $%
\mu $, and functionals of the external field $\phi (\mathbf{r})$ and of the
(pair and any higher order) intermolecular potentials. Here $\Xi $ is the
Grand partition function and we will let $\beta =(k_{B}T)^{-1}$. We are
particularly interested in the functional dependence on the external field $%
\phi (\mathbf{r})$ for fixed values of the thermodynamic variables and the
intermolecular potentials. Since the zero of energy can be chosen
arbitrarily without affecting the physics, and a constant field acts like a
chemical potential shift in the Grand partition function, the relevant
quantities actually depend only on the difference \cite{3} 
\begin{equation}
\Phi (\mathbf{r})\equiv \phi (\mathbf{r})-\mu  \label{Phi1}
\end{equation}
as shown in the Appendix. We refer to $\Phi (\mathbf{r})$ as the
\emph{singlet field} and write $\Omega ([\Phi ])$ or $\rho (\mathbf{r;}[\Phi ])$
when we want to emphasize this functional dependence.

In this case the Gibbs inequality can be written as: 
\begin{equation}
\Omega _{t}+\langle H(\Phi )-H_{t}(\Phi _{t})\rangle \,\leq \,\Omega \,\leq
\,\Omega _{t}+\langle H(\Phi )-H_{t}(\Phi _{t})\rangle _{t}.
\label{generalupperlowerbound}
\end{equation}
Here $\langle \; \rangle $ and $\langle \; \rangle _{t}$ denote normalized
ensemble averages in the systems with Hamiltonians $H(\Phi )$ and $%
H_{t}(\Phi _{t})$ respectively, and $\Omega $ and $\Omega _{t}$ are the
respective free energies. This result follows immediately from Eq.~(4) of
the basic paper by Mermin \cite{6}; it holds for general quantum systems
and, with modified definitions of the free energies and averages, for other
ensembles too. For completeness and to establish notation, we give in the
Appendix a derivation for classical systems that uses only convexity
properties of the exponential function and seems a little simpler than
standard approaches \cite{2,4,5,10}. As emphasized in the Appendix, the
inequalities in Eq.~(\ref{generalupperlowerbound}) are strict; only if $%
H(\Phi )=H_{t}(\Phi _{t})$ for \emph{every configuration} with nonzero
weight in the averages will $\Omega _{t}=\,\Omega .$

\section{Density response to different singlet fields}

Consider first the special case where the trial system has exactly the same
intermolecular interactions as the system of interest and only the singlet
fields may differ. By combining the lower and upper bounds in Eq.~(\ref
{generalupperlowerbound}) for this case we find a simple inequality 
\begin{equation}
0\,\leq \int d\mathbf{r}[\Phi (\mathbf{r})-\Phi _{t}(\mathbf{r})][\rho _{t}
(\mathbf{r})-\rho (\mathbf{r})]\,.  \label{singletdensitybound}
\end{equation}
Here $\rho _{t}(\mathbf{r})\,\mathbf{\equiv \,}\rho (\mathbf{r;[}\Phi _{t}%
\mathbf{])}$ is the equilibrium singlet density arising from $\Phi _{t}$ and
similarly $\rho (\mathbf{r})\,\mathbf{\equiv \,}\rho (\mathbf{r;[}\Phi 
\mathbf{]).}$ Equation (\ref{singletdensitybound}) is valid for any $\Phi
_{t}(\mathbf{r})$. Since the inequalities in Eq.~(\ref{generalupperlowerbound})
are strict, the right side of Eq. (\ref{singletdensitybound}) is strictly
greater than zero unless $\Phi _{t}(\mathbf{r})=\Phi (%
\mathbf{r})$ for all $\mathbf{r.}$

Equation (\ref{singletdensitybound}) has some interesting consequences for
the density response of a system to an external field. In particular, it
shows \emph{explicitly} that the mapping from the singlet field
$\Phi (\mathbf{r)}$
to the associated equilibrium density $\rho (\mathbf{r;[}\Phi \mathbf{])}$
is one to one: a given singlet field produces a \emph{unique} density
response. Thus if $\Phi _{t}(\mathbf{r})$ differs from $\Phi (\mathbf{r)}$
for any $\mathbf{r}$ then the right side of Eq.~(\ref{singletdensitybound})
is strictly greater than zero; for this to be true $\rho _{t}(\mathbf{r})$
must differ from $\rho (\mathbf{r})$ for some $\mathbf{r.}$ This result
plays a central role in density functional theory and is well known. Mermin
proved this for quantum systems by contradiction \cite{6}, and most later
discussions of density functional theory \cite{2,3,4,5} for classical
systems have also relied on proofs by contradiction \cite{10a}.

However Eq.~(\ref{singletdensitybound}) allows us very simply to find more
explicit results in some special cases that do not appear to be widely
known. For example, suppose $\Phi _{t}(\mathbf{r})$ differs from $\Phi (%
\mathbf{r} ) $ only in a finite region of space $\Gamma ,$ which can be
arbitrarily small. Then the integration in Eq.~(\ref{generalupperlowerbound}%
) is restricted to $\Gamma $ and we find the more local result that $\rho
_{t}(\mathbf{r})$ must differ from $\rho (\mathbf{r})$ for some $\mathbf{r}$
in $\Gamma .$ (This approach gives no information about what happens outside 
$\Gamma .)\;$This result is obvious in the low density limit where there is
a local density response proportional to the Boltzman factor of $\Phi (%
\mathbf{r})$ but it holds true in general. If in addition $\Phi _{t}(\mathbf{%
r})>\Phi (\mathbf{r})$ in $\Gamma $ then $\int_{\Gamma }d\mathbf{r}[\rho (%
\mathbf{r})-\rho _{t}(\mathbf{r})]>0$. More generally, if $\Gamma $ denotes
the set of points where $\Phi _{t}(\mathbf{r})$ differs from $\Phi (\mathbf{r%
}),$ then $\rho _{t}(\mathbf{r})$ must differ from $\rho (\mathbf{r})$ at
least for some $\mathbf{r}$ in $\Gamma .$

One well known result follows when $\Delta \Phi _{t}(\mathbf{r})$ $\mathbf{%
\equiv } $ $\Phi _{t}(\mathbf{r})-\Phi (\mathbf{r})$ is very small, as
indicated by the notation $\delta \Phi _{t}(\mathbf{r})$. Then we can use
linear response theory to determine the small induced density change: 
\begin{equation}
\delta \rho _{t}(\mathbf{r})=-\beta \int d\mathbf{r}^{\prime }\chi (\mathbf{r%
},\mathbf{r}^{\prime };[\Phi ])\delta \Phi _{t}(\mathbf{r}^{\prime }),
\label{linearresponse}
\end{equation}
and Eq.~(\ref{singletdensitybound}) reduces to 
\begin{equation}
0\,<\int d\mathbf{r}d\mathbf{r}^{\prime }\delta \Phi _{t}(\mathbf{r})\chi (%
\mathbf{r},\mathbf{r}^{\prime };[\Phi ])\delta \Phi _{t}(\mathbf{r}^{\prime
}),  \label{chipositive}
\end{equation}
which expresses that fact that the linear response function $\chi $ is
positive definite \cite{4,5} (except possibly at a phase transition in zero
field where our assumption that linear response theory is accurate for
sufficiently small $\delta \Phi _{t}(\mathbf{r})$ can break down).

\section{Density functional theory}

The key idea in density functional theory \cite{2,3,4,5,6,7} is to consider
generalized free
energies that are functionals of the density $\rho _{t}(\mathbf{r)}$ rather
than the field $\Phi _{t}(\mathbf{r})$. Since the Gibbs inequality shows
that a given singlet field produces a unique density response as discussed
above, formally any functional of $\Phi _{t}$ can equally well be thought of
as a functional of $\rho _{t}.$ But the Gibbs inequality can play a more
central role. We will show that the basic functionals and the minimum
principle used in density functional theory arise directly from the Gibbs
inequality and its use permits us to see explicitly how the change of
variable from singlet fields to densities can be carried out.

We start from the upper bound to the free energy $\Omega $ that follows from
Eq.~(\ref{generalupperlowerbound}) in the special case where only the
singlet fields differ:
\begin{equation}
\Omega \,\leq \,\Omega _{t}+\int d\mathbf{r}[\Phi (\mathbf{r})-\Phi _{t}(%
\mathbf{r})]\rho _{t}(\mathbf{r})\,.  \label{singletupper}
\end{equation}
The standard density functional relations \cite{2,3,4,5} follow immediately
by simply rearranging and reinterpreting terms in this equation. Thus we
rewrite Eq.~(\ref{singletupper}) as 
\begin{equation}
\Omega =\min_{\Phi _{t}}\left[ \left\{ \Omega _{t}-\int d\mathbf{r}\rho _{t}(%
\mathbf{r})\Phi _{t}(\mathbf{r})\right\} +\int d\mathbf{r}\rho _{t}(\mathbf{r%
})\Phi (\mathbf{r})\right];  \label{minphi}
\end{equation}
here the minimum is taken over all possible fields $\Phi _{t}$ and is
achieved only when $\Phi _{t}=\Phi $ for all $\mathbf{r}$.

Equation (\ref{minphi}) may seem to be a particularly hard and inefficient
way to calculate $\Omega $ but we can proceed as follows. Since the density
is related to the functional derivative of the free energy by 
\begin{equation}
\rho _{t}(\mathbf{r})\equiv \rho (\mathbf{r;[}\Phi _{t}])=\delta \Omega ([\Phi
_{t}])/\delta \Phi _{t}(\mathbf{r}),  \label{domegadphi}
\end{equation}
the terms in curly brackets in the right side of Eq.~(\ref{minphi})
represent a (functional) Legendre transform \cite{7} from $\Omega _{t}\equiv 
\Omega ([\Phi _{t}])$ to a so-called \emph{intrinsic free energy density
functional} $F([\rho _{t}]),$ where 
\begin{equation}
F([\rho _{t}])\equiv \Omega ([\Phi _{t}])-\int d\mathbf{r}\rho _{t}(\mathbf{%
r)}\Phi _{t}(\mathbf{r})  \label{Flegendre}
\end{equation}
denotes the terms in curly brackets. As the notation indicates, $F$ is a
functional of the equilibrium density $\rho _{t}(\mathbf{r)}$ and not of the
field; thus the $\Phi _{t}(\mathbf{r})$ on the right side is the field that
corresponds to the density $\rho _{t}(\mathbf{r)}$. By standard properties
of the Legendre transform the associated field formally satisfies 
\begin{equation}
\delta F([\rho _{t}])/\delta \rho _{t}(\mathbf{r})=-\Phi _{t}(\mathbf{r}).
\label{dFdrho}
\end{equation}

The transformation of variables from fields to densities in $F$ can be seen
more explicitly if we rewrite Eq.~(\ref{Flegendre}) in the following
equivalent form: 
\begin{equation}
F([\rho _{t}])\equiv \,\min_{\Phi _{t}^{\prime }}\left[ \Omega ([\Phi
_{t}^{\prime }])-\int d\mathbf{r}\rho _{t}(\mathbf{r)}\Phi _{t}^{\prime }(%
\mathbf{r} )\right].  \label{Flegendremin}
\end{equation}
The minimum is taken over all possible external fields $\Phi _{t}^{\prime }$
at fixed $\rho _{t}(\mathbf{r)}$. The minimum occurs when the functional
derivative with respect to $\Phi _{t}^{\prime }$ of the term in square
brackets vanishes; from Eq.~(\ref{domegadphi}) this occurs when $\Phi
_{t}^{\prime }$ equals the field $\Phi _{t}$ associated with the density
distribution $\rho _{t}(\mathbf{r)}$, as in Eq.~(\ref{Flegendre}). Thus $%
F([\rho _{t}])$ is certainly well-defined in principle, though Eq.~(\ref
{Flegendremin}) does not present a very useful way to calculate it in
practice.

Having made this change of variable, the only remaining dependence on the
field $\Phi _{t}$ on the right side of Eq.~(\ref{minphi}) is through the
associated equilibrium density $\rho _{t}(\mathbf{r),}$ and the minimum can
be taken over all such density distributions. Thus Eq.~(\ref{minphi}) can be
rewritten as 
\begin{equation}
\Omega =\Omega ([\Phi ])=\min_{\rho _{t}}\left[ F([\rho _{t}])+\int d\mathbf{%
r}\rho _{t}(\mathbf{r})\Phi (\mathbf{r})\right].  \label{minrho}
\end{equation}

This is the basic equation of density functional theory. The term in square
brackets on the right side is usually interpreted as a generalized \emph{%
Grand free energy density functional} $\Omega _{\Phi }([\rho _{t}])$ \cite{3}:
\begin{equation}
\Omega _{\Phi }([\rho _{t}])\equiv F([\rho _{t}])+\int d\mathbf{r}\rho _{t}(%
\mathbf{r)}\Phi (\mathbf{r}).  \label{granddf}
\end{equation}
Note that $\Omega _{\Phi }([\rho _{t}]),$ unlike $F([\rho _{t}])$, is a
functional of the field $\Phi $ as well as the density $\rho _{t}.$ (Of
course, both $\Omega _{\Phi }$ and $F$ remain functionals of all the
intermolecular interaction potentials). Equation (\ref{minrho}) shows
explicitly that the minimum value of $\Omega _{\Phi }([\rho _{t}])$ with
respect to $\rho _{t}$ at fixed $\Phi $ is the desired free energy $\Omega %
([\Phi ]).$ The density that minimizes this functional is $\rho (\mathbf{r})$
and at that minimum we have 
\begin{equation}
\frac{\delta \Omega _{\Phi }([\rho _{t}])}{\delta \rho _{t}(\mathbf{r})}=0.
\label{granddfmin}
\end{equation}

This derivation highlights the following points. The minimum in Eq.~(\ref
{minrho}) must be taken over equilibrium densities $\rho _{t}$ arising from
a system with the same intermolecular interactions but a different singlet
field. While the mapping from arbitrary fields to equilibrium densities is
one to one, it is not necessarily true that an arbitrary function of
the density must correspond to the equilibrium density of a system with a
given $H$ in some field (unless the field is trivially taken to be so much
stronger than the intermolecular interactions that there is an essentially
local response). Thus, a density distribution predicting 14 nearest
neighbors about a fixed particle in a hard sphere system is unphysical and
should not be included in the trial density set for Eq.~(\ref{minrho}). In
practice, we suspect this is not a serious
problem in most cases where only limited density variations are considered
and for non hard core systems the inverse mapping can generally be carried
out \cite{11}.

Analogous equations hold in the Canonical ensemble, but then the trial
density set should include only densities yielding the fixed number of
particles. In this case workers have found it more convenient to work in the
Grand ensemble where this restriction does not apply and then subtract out
the lowest order effects of particle number fluctuations \cite{12}.

\section{Different intermolecular interactions}

More problematic is the fact that the functional dependence on $\rho _{t}$
is also affected by the intermolecular interactions. While the exact $%
F([\rho _{t}])$ is a ``universal'' functional of $\rho _{t}$ for a given set
of intermolecular interactions, it remains an essentially unknown functional
of the intermolecular interactions, which in classical applications can have
a variety of forms. Any approximate expression for $F([\rho _{t}])$ must
implicitly or explicitly take this functional dependence into account. Thus
a density functional that can accurately describes the properties of a
nonuniform hard sphere fluid may not be appropriate for a system with softer
repulsive intermolecular interactions. This problem seems much more severe
when there are attractive intermolecular interactions. Then $F$ must
describe critical properties, capillary waves, and a variety other phenomena
where the functional dependence on $w$ plays an essential role, and we have
little idea of its form \cite{13}. (And of course knowing the exact form, as
we do for the original free energy $\Omega ([\Phi ])$ through the partition
function, does not mean we can carry out practical calculations!)

In many cases where long wavelength fluctuations are not important, a mean
field treatment of the attractive interactions may suffice. For the simple
Lennard-Jones (LJ) fluid with pair interactions $w_{LJ}(r_{12})$ this can
arise by approximating the structure of the nonuniform LJ fluid by that of a
\emph{reference fluid} with purely repulsive pair interactions $u_{0}(r_{12})$
giving repulsive forces equal to those of the LJ potential and a singlet
field $\Phi _{0}(\mathbf{r})$ that incorporates the averaged effects of the
attractive interactions \cite{13}. The free energy can again be estimated
using the Gibbs inequality, but now in a more general form than Eq.~(\ref
{singletupper}) or (\ref{minrho}), since the intermolecular pair potentials
as well as the singlet fields differ. We use the subscript $0$ to denote the
trial system in this special case. Equation (\ref{generalupperlowerbound})
then becomes
\begin{equation}
\begin{array}[b]{ll}
\Omega _{t}-\int d{\bf r}\rho ({\bf r)}\Delta \Phi ({\bf r})+\frac{1}{2}\int
d{\bf r}_{1}d{\bf r}_{2}\rho ^{(2)}({\bf r}_{1}{\bf ,r}_{2}{\bf )}%
u_{1}(r_{12}) & \leq  \\ 
\;\;\;\;\;\;\;\;\;\;\;\;\Omega \;\leq \,\,\Omega _{t}-\int d{\bf r}\rho _{0}(%
{\bf r})\Delta \Phi ({\bf r}) &  \\ 
\;\;\;\;\;\;\;\;\;\;\;\;\;\;\;\;\;\;\;\;\;\;+\frac{1}{2}\int 
d{\bf r}_{1}d{\bf r}_{2}\rho
_{0}^{(2)}({\bf r}_{1}{\bf ,r}_{2}{\bf )}u_{1}(r_{12}) \,. & 
\end{array}
\label{pairupperlower}
\end{equation}
%\begin{equation}
%\begin{array}{ll}
%\Omega _{t}-\int d\mathbf{r}\rho (\mathbf{r)}\Delta \Phi (\mathbf{r})+\frac{1%
%}{2}\int d\mathbf{r}_{1}d\mathbf{r}_{2}\rho ^{(2)}(\mathbf{r}_{1}\mathbf{,r}%
%_{2}\mathbf{)}u_{1}(r_{12}) & \leq \,\Omega \;\leq  \\ 
%\Omega _{t}-\int d\mathbf{r}\rho _{0}(\mathbf{r})\Delta \Phi (\mathbf{r})+%
%\frac{1}{2}\int d\mathbf{r}_{1}d\mathbf{r}_{2}\rho _{0}^{(2)}(\mathbf{r}_{1}%
%\mathbf{,r}_{2}\mathbf{)}u_{1}(r_{12}) & 
%\end{array}
%\label{pairupperlower}
%\end{equation}
Here $u_{1}(r_{12})=w_{LJ}(r_{12})-u_{0}(r_{12}).$

To get an upper bound to
the free energy from the last inequality, we need an accurate expression for
the pair distribution function $\rho _{0}^{(2)}(\mathbf{r}_{1}\mathbf{,r}_{2}%
\mathbf{)}$ as well as $\rho _{0}(\mathbf{r).}$ However, since $u_{0}(r_{12})
$ is different than $w_{LJ}(r_{12})$ the Gibbs inequality shows there is no
choice of $\Phi _{0}(\mathbf{r})$ for which the upper bound gives the exact
free energy. Nevertheless, if $\Phi _{0}(\mathbf{r})$ is chosen so that $%
\rho _{0}(\mathbf{r})$ closely approximates $\rho (\mathbf{r)}$, it seems
physically plausible that $\rho _{0}^{(2)}(\mathbf{r}_{1}\mathbf{,r}_{2}%
\mathbf{)}$ could also rather closely resemble $\rho ^{(2)}(\mathbf{r}_{1}%
\mathbf{,r}_{2}\mathbf{),}$ at least at high density when correlations
arising from packing of the repulsive cores are most important \cite{13}.
When this is true, the upper and lower bounds in Eq.~(\ref{pairupperlower})
are close to each other and the upper bound provides a reasonable estimate
for the exact free energy. The basic equations of perturbation theory of \emph{%
uniform} fluids can be established from this perspective and accurate
results for the free energy of the dense LJ uniform fluid are found \cite{14}
from the upper bound in Eq.~(\ref{pairupperlower}).

An upper bound for the free energy from the last inequality is guaranteed
only if we use accurate values for $\rho _{0}(\mathbf{r})$ and $\rho
_{0}^{(2)}(\mathbf{r}_{1}\mathbf{,r}_{2}\mathbf{).}$ In many applications of
density functional theory $\rho _{0}^{(2)}(\mathbf{r}_{1}\mathbf{,r}_{2}%
\mathbf{)}$ is replaced by the product $\rho _{0}(\mathbf{r}_{1})\rho _{0}(%
\mathbf{r}_{2})$ for computational simplicity \cite{3}. A minimization of
the approximate free energy when this approximation is made cannot be
justified by any upper bound principle, though it may be useful for other
purposes. In a uniform fluid this approach yields a
constant value for $\Phi _{0}(\mathbf{r})$, which gives a constant reference
density equal to the LJ density, and the resulting free energy is
only qualitatively accurate, sometimes being larger and sometimes smaller
than the correct result \cite{13,15}.

\section{Acknowledgments}

It is a pleasure to dedicate this paper to Michael Fisher on the happy
occasion of his 70th birthday and to thank him for his unfailing support,
insight, and advice over the many years we have known each other. We are also
grateful to Yng-Gwei Chen and Ted Kirkpatrick for helpful discussions. This work
is supported by the National Science Foundation through Grant CHE-0111104.

\appendix
\section{}
Consider a classical system with interaction potentials 
\begin{equation}
H(\phi )=\sum\limits_{i=1}^{N}\phi (\mathbf{r}_{i})+\sum%
\limits_{i<j}^{N}w^{(2)}(\mathbf{r}_{i},\mathbf{r}_{j})+ \cdot \cdot \cdot
\label{interactionpotentials}
\end{equation}
We explicitly denote only the dependence of $H$ on the external field $\phi ;
$ the intermolecular interactions $w^{(j)}$ are general and can include $%
3,4,...,N$ body terms. The Grand partition function $\Xi $ for a system with
chemical potential $\mu ,$ and temperature $k_{B}T\equiv \beta ^{-1}$ is given
by
\begin{equation}
\Xi \equiv e^{-\beta \Omega }=\mathrm{Tr}\,e^{-\beta H(\phi )+\beta \mu N}=%
\mathrm{Tr}\,e^{-\beta H(\Phi )},  \label{grandpartition}
\end{equation}
where 
\begin{equation}
\mathrm{Tr}\,(\cdot )\equiv \sum\limits_{N}[N!\Lambda ^{3N}]^{-1}\int d\mathbf{r}%
^{N}(\cdot )\,.  \label{trace}
\end{equation}
Here $\Lambda \equiv (\beta h^{2}/2\pi m)^{1/2}$ is the de Broglie
wavelength. In the last equality in Eq.~(\ref{grandpartition}) we combined the
chemical potential terms with the external field terms in Eq.~(\ref
{interactionpotentials}) to exhibit the functional dependence only on the
the singlet field $\Phi (\mathbf{r})\equiv \phi (\mathbf{r})-\mu $.
Finally we use pointed brackets to
define a normalized ensemble average: 
\begin{equation}
\langle (\cdot )\rangle \equiv e^{\beta \Omega }\mathrm{Tr}
e^{-\beta H(\Phi )}(\cdot )\,.  \label{normalizedaverage}
\end{equation}

To arrive at the Gibbs inequality, we consider an arbitrary ``trial'' system
with different interaction potentials $H_{t}(\Phi _{t})$ and note that the
partition function $\Xi =e^{-\beta \Omega }$ can be written as 
\begin{equation}
e^{-\beta \Omega }=\mathrm{Tr}e^{-\beta H_{t}(\Phi
_{t})}e^{-\beta [H(\Phi )-H_{t}(\Phi _{t})]}  \label{exact1}
\end{equation}
or 
\begin{equation}
e^{-\beta \Omega }=e^{-\beta \Omega _{t}}\langle e^{-\beta [H(\Phi
)-H_{t}(\Phi _{t})]}\rangle _{t}\,. \label{exact2}
\end{equation}
If $x$ is a random variable and $\langle x\rangle $ denotes its average over
any normalized probability distribution we have $e^{x-\langle x\rangle }%
\geq 1+x-\langle x\rangle $ for all real $x$. Taking averages we find the
familiar result $\langle e^{x}\rangle \geq e^{\langle x\rangle }$ with the
strict inequality holding if there is any configuration in the average with
$x\neq \langle x\rangle $. Applying this to Eq.~(\ref{exact2}) gives 
\begin{equation}
e^{-\beta \Omega }\,\geq \,e^{-\beta \Omega _{t}}e^{-\beta \langle [H(\Phi
)-H_{t}(\Phi _{t})]\rangle _{t}}\,.  \label{partitioninequality}
\end{equation}
This yields the final result 
\begin{equation}
\Omega \,\leq \,\Omega _{t}+\langle H(\Phi )-H_{t}(\Phi _{t})\rangle _{t},
\label{gibbsupper}
\end{equation}
where the equality holds only when $H(\Phi )=H_{t}(\Phi _{t})$ for all
configurations in the average. Swapping $H(\Phi )$ and $H_{t}(\Phi _{t})$
yields the lower bound as given in Eq.~(\ref{generalupperlowerbound}). By
redefining the averages and free energies appropriately these results also
hold in the Canonical ensemble, and they are valid for
quantum systems as well \cite{6}.

\end{document}